\documentclass[twocolumn,aps,prl,showpacs,superscriptaddress,twocolumn]{revtex4-1}
\usepackage{lmodern}

\usepackage[latin9]{inputenc}
\usepackage{amsmath}
\usepackage{amssymb}
\usepackage{graphicx}
\usepackage{epstopdf}
\usepackage{color}
\usepackage{enumerate}
\usepackage{ulem}
\usepackage{natbib}
\usepackage{caption}
\usepackage{soul,color}
\definecolor{orange}{RGB}{255,125,0}

\begin{document}
	
	\bibliographystyle{apsrev}
	
	\title{The Temperature dependence of the magneto-resistance peak in highly disordered superconductors}

	\author{A. Doron}
	\email{adam.doron@weizmann.ac.il; Corresponding author}
	\affiliation{Department of Condensed Matter Physics, The Weizmann Institute of Science, Rehovot 7610001, Israel.}
	
	\author{I. Tamir}
	\affiliation{Department of Condensed Matter Physics, The Weizmann Institute of Science, Rehovot 7610001, Israel.}
	
	\author{T. Levinson}
	\affiliation{Department of Condensed Matter Physics, The Weizmann Institute of Science, Rehovot 7610001, Israel.}

	\author{F. Gorniaczyk}
	\affiliation{Department of Condensed Matter Physics, The Weizmann Institute of Science, Rehovot 7610001, Israel.}	

	\author{D. Shahar}
	\affiliation{Department of Condensed Matter Physics, The Weizmann Institute of Science, Rehovot 7610001, Israel.}

	\begin{abstract}
Highly disordered superconductors have a rich phase diagram. At a moderate magnetic field ($B$) the samples go through the superconductor-insulator quantum phase transition. In the insulating phase, the resistance increases sharply with $B$ up to a magneto-resistance peak beyond which the resistance drops with $B$.
In this manuscript we follow the temperature ($T$) evolution of this magneto-resistance peak. We show that as $T$ is reduced, the peak appears at lower $B$'s approaching the critical field of the superconductor-insulator transition. 
Due to experimental limitations we are unable to determine whether the $T=0$ limiting position of the peak matches that of the critical field or is at comparable but slightly higher $B$.
We show that, although the peak appears at different $B$ values, its resistance follows an activated $T$ dependence over a large $T$ range with a prefactor that is very similar to the quantum of resistance for cooper-pairs.

	\end{abstract}
	
	\maketitle	


Highly disordered superconductors undergo a superconductor-insulator quantum phase transition (SIT) \cite{goldmanpt51,physupekhi,sondhirmp} driven by experimentally tunable parameters such as $B$ \cite{HebardPrl,kapitulnikprl74,BaturinaJETP}, disorder strength \cite{Shaharprb}, carrier density \cite{goldmanprl94} or sample thickness \cite{haviprl62,goldmanpt51}.

The $B$ driven insulating phase, which emerges above a critical $B$ ($B_{c}$), exhibits a pronounced peak in the magnetoresistance (MR) \cite{paalanenprl69}. $R(T)$ measured at relaltively high $T$'s in the MR peak has a characteristic activation $T$ similar to the $B=0$ superconducting $T_{c}$ \cite{murthyprl2004}. Therefore the MR peak is typically associated with a state where cooper pairs persist above $B_{c}$ but become spatially localized. This view of a Cooper-pair insulator is supported by several theoretical and experimental studies \cite{FeigAnnals,YonatanNat,GantmakherJETP,vallesprl103,BenjaminNat,sacepe2015high,craneprb751} and some works consider the MR peak itself as the transition point (or crossover) between a bosonic insulator closely above $B_c$ and a fermionic insulator at high $B$'s \cite{paalanenprl69,breznay2016self,steiner2005superconductivity}.

The $B$ driven insulating phase in our system can be separated into two distinguishable regions that show different $B$ and $T$ dependences. To distinguish between these regimes we first write the $R(T)$ dependence of our insulator as
\begin{equation}
	R(T)=R_{0}exp(T_{0}/T)^{\gamma}
	\label{eTdependence}
\end{equation}
where $R_{0}$ is a constant, $k_{B}T_{0}$ is the energy characterizing the conduction process and $\gamma\in[0,1]$ (when $\gamma=1$ the conduction is termed activated).
In our system, at $B$'s slightly above $B_{c}$, $R$ increases with increasing $B$ and $R$ has either an activated $T$ dependence (for $1$K$ >T>200$mK) or a novel $T$ dependence where $R$ seems to diverge at a finite $T$ ($T^{*}$) \cite{MaozFiniteT} (the $R(T)$ follows a phenomenological fit similar to equation (\ref{eTdependence}) but with $T\to (T-T^{*})$).
The second transport regime appears at high $B$'s (typically above $6$ T) where $R$ has a sub-activated $T$ dependence ($\gamma<1$) and, at a constant $T$, $R$ decreases with increasing $B$. The transition between these two regimes occurs in the vicinity of the MR peak.

In this work we provide a systematic investigation of the $T$-dependence of the MR peak.
We show that, at low $T$'s, the $B$ where the MR peak appears ($B_{peak}$) decreases rapidly. 
Extrapolating our lower-$T$ $B_{peak}$ raises the possibility that in the limit $T\to 0$, $B_{peak}\to B_{c}$ \cite{Footnote1}.
Furthermore, studying the $R$ values at the MR peak we show that, although measured at different $B$'s, $R(B_{peak})$ has an activated $T$-dependence over a wide $T$ range. The prefactor of this activation is near the quantum of resistance for Cooper-pairs.

For this study we used data obtained from 23 different thin films of highly disordered amorphous indium oxide ($\alpha$-InO).
The samples were deposited by e-gun evaporation of high purity In$_2$O$_3$ pellets onto a Si/SiO$_2$ substrate in an Oxygen rich environment (typically $1.5\cdot 10^{-5}$ Torr). 
Most samples were hall-bar shaped and the thickness of samples, measured while evaporating using a crystal monitor, ranged between 28 and 40 nm.
The contacts of samples were either Ti/Au contacts prepared via optical lithography prior to the In$_2$O$_3$ evaporation or pressed indium.

\begin{figure} [h!]
		\includegraphics [width=8.5 cm] {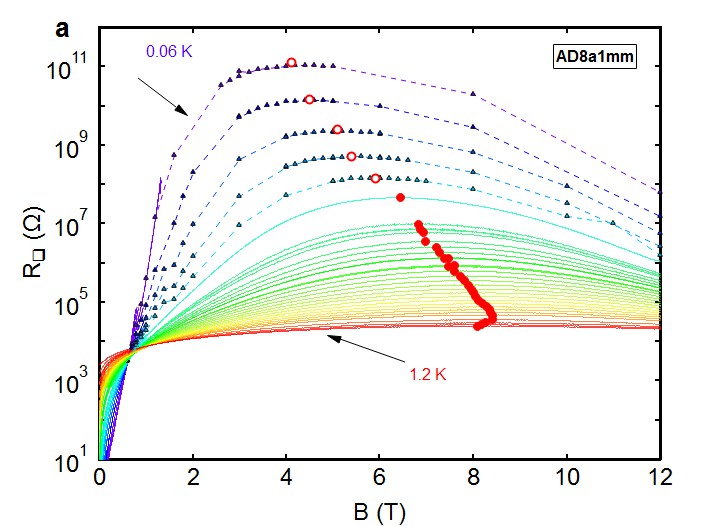}
		\includegraphics [width=8.5 cm] {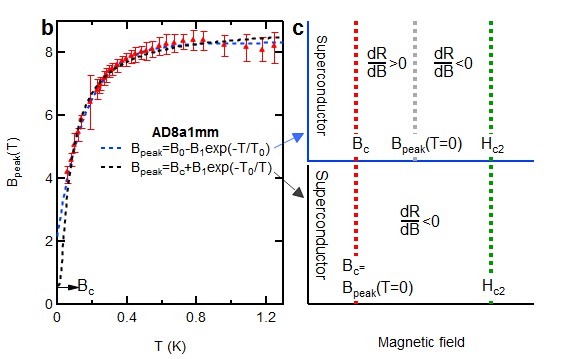}

		\caption{{\bf T-evolution of the MR Peak.} 
			(a) $R$ (log scale) vs $B$ of sample AD8a1mm. The color-coding stands for different isotherms ranging from 20 mK (purple) to 1.2 K (red). Full lines correspond to data extracted via 2 and 4 terminal Lock-in measurements. Triangles connected via dashed lines correspond to data extracted from the DC $I-V$ characteristics. The red circles mark the MR peak at each T.
			(b) $B_{Peak}$ vs $T$. The red triangles are $B_{peak}$ extracted from the data of figure (a). The dashed lines are fits to different phenomenological functional forms, the main difference between these forms is that according to the blue line $B_{peak}^{0}=2.16 T>B_{c}$ and according to the black line $B_{peak}^{0}=B_{c}$.
			(c) Two possible $T\to 0$ phase diagrams. The top, blue, (bottom, black,) diagram corresponds to the blue (black) fit in (b). The main difference between the phase diagrams is that the bottom diagram predicts that $\frac{dR}{dB}(T\to 0)<0$ throughout the insulating phase.
			}			
		\label{Figure1}
	\end{figure}

In figure \ref{Figure1}a we follow the $T$ dependence of the insulating peak by displaying $R$ vs $B$ of sample AD8a1mm at $T\in(0.06,1.2)$K. The continuous lines were measured by 4 probe ($R<100 K\Omega$ data, excitation current of 0.1 nA) and 2 probe ($R>100 K\Omega$, excitation voltage of 10$\mu$Volt) techniques. 
The black triangles connected via dashed lines mark $R$ extracted from 2 probe dc current-voltage characteristics. 
The red circles mark the MR peak of each isotherm (see supplemental material section S2 for details of our peak detection algorithm). 
The difference between full and empty circles will be explained in a following section. 
Upon decreasing $T$, $B_{peak}$ increases slightly down to $0.6$ K and then decreases rapidly.

This drop in $B_{peak}$ is puzzling because, as mentioned above, the MR peak is associated with the process of termination of localized Cooper-pairs \cite{paalanenprl69,breznay2016self,steiner2005superconductivity}. As the superconducting order parameter, which is non-zero locally in a Cooper-pair insulator, diminishes with increasing $T$ one would expect that at low $T$'s local superconductivity will withstand higher $B$'s therefore one would expect $B_{peak}$ to increase at low $T$'s and approach $H_{c2}\sim 14$T \cite{sacepe2015high} as $T\to0$. 
Our results show otherwise, $B_{peak}$ decreases significantly as $T\to 0$. 

In figure \ref{Figure1}b we display $B_{peak}$ vs $T$. The error bars mark our uncertainty interval of $B_{peak}$ (in section S2 of the supplemental material we elaborate on our error determination procedure).
In order to study the ground state of the system and the $T=0$ quantum phase transition it is important to know what is the $T=0$ limit of the peak, $B_{peak}^{0}\equiv B_{peak}(T\to 0)$. 
Because $B_{peak}^{0}$ in highly disordered samples requires measuring diverging $R$'s it is not experimentally measurable. 
In the absence of a relevant theory we have to fit the $B_{peak}^{0}$ data to a phenomenological form and extrapolate it to $T=0$. 
The dashed lines in figure \ref{Figure1}b are two such fits. 
From the first fit (blue line) we get $B_{peak}(T=0)=2.16T>B_{c}(T=0)=0.55T$ according to which even as $T\to 0$ there is a MR peak at $B>B_{c}$. 
The dashed black line is the other phenomenological fit where we placed the constraint $B_{peak}^{0}=B_{c}$. 
Since both fits are within the errors in $B_{peak}(T)$ we cannot rule out the possibility that $B_{peak}^{0}=B_{c}$. 
We note that $B_{c}$ can also have a weak $T$ dependence, more on this in the discussion section.

In figure \ref{Figure1}c we sketch two possible $T=0$ phase diagrams, the top (bottom) phase diagram corresponds to the black (blue) fit in figure \ref{Figure1}b.
For the best of our knowledge the possibility that $B_{peak}^{0}=B_{c}$ (bottom phase diagram) is raised here for the first time. 
In the discussion section we present a third possible phase diagram that arises due to the phenomenon of the finite-$T$ insulator \cite{MaozFiniteT}.

\begin{figure} [h!]
		\includegraphics [width=8.5 cm] {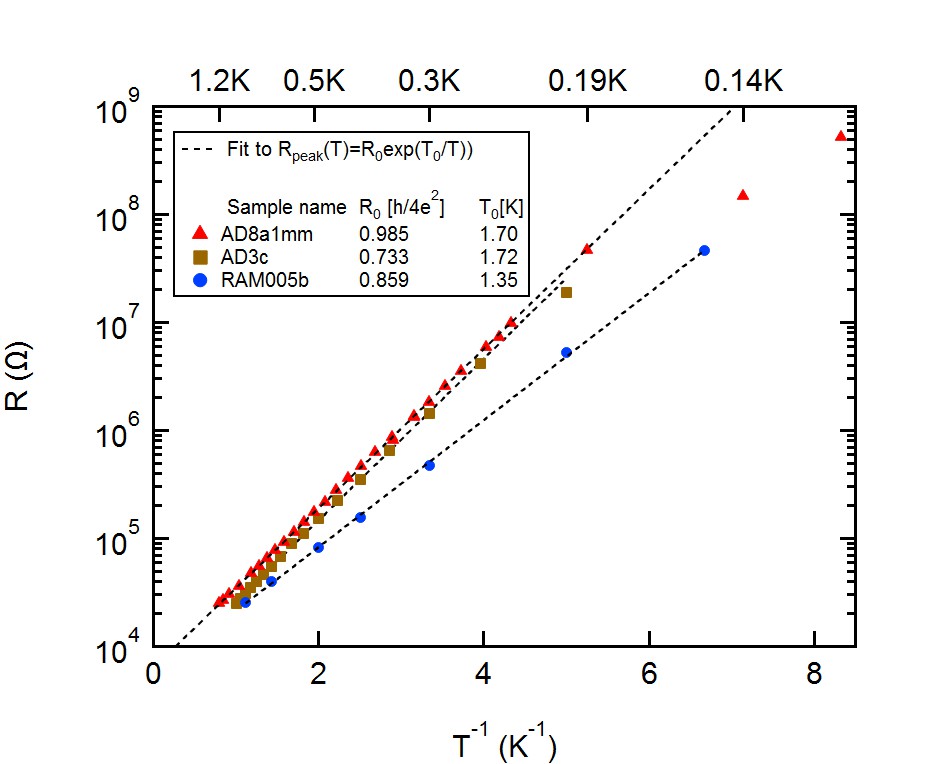}
		
		\caption{{\bf T-Activation of MR peak.} 
		$R_{peak}$ (log scale) vs $T^{-1}$ of three samples where the red triangles were extracted from the data of figure \ref{Figure1}a. The dashed black lines are activated fits to $R_{Peak}(T)=R_{0}exp(T_{0}/T)$. For sample AD8a1mm we get $R_{0}=0.985 \frac{h}{4e^{2}}$ and $T_{0}=1.7$K where the fit holds for $T\in (0.19,1.2)$K below which $R_{Peak}$ is sub-activated.
		}			
		\label{Figure2}
	\end{figure}

It is interesting to consider the $R(T)$ dependence at the MR peak ($R_{peak}$) itself. In figure \ref{Figure2} we display $R_{peak}$ (log scale) vs $T^{-1}$ of three different samples. The dashed black lines are fits to equation (\ref{eTdependence}) with $\gamma =1$.
Although at every $T$ $R_{peak}$ is measured at a different $B$ (see figures \ref{Figure1}a,b), it can be seen that overall it has an activated behavior between 0.19 and 1.2 K (for sample AD8a1mm). 
In section S1 of the supplemental material we compare between the activated behavior of $R(B_{peak})$ and $R(B)$ for all constant $B$'s and show that the fit quality of the peak is better than that of any other constant $B$.
At $T<$0.19 K $R_{peak}$ deviates from the activated behavior (these points are marked in figure \ref{Figure1}a by empty circles). We discuss this low-$T$ deviation in the discussion section.
In section S3 of the supplemental material we plot $R_{peak}$ vs $T^{-1}$ for 18 different samples that showed an activated peak.

\begin{figure} [h!]
		\includegraphics [width=8.5 cm] {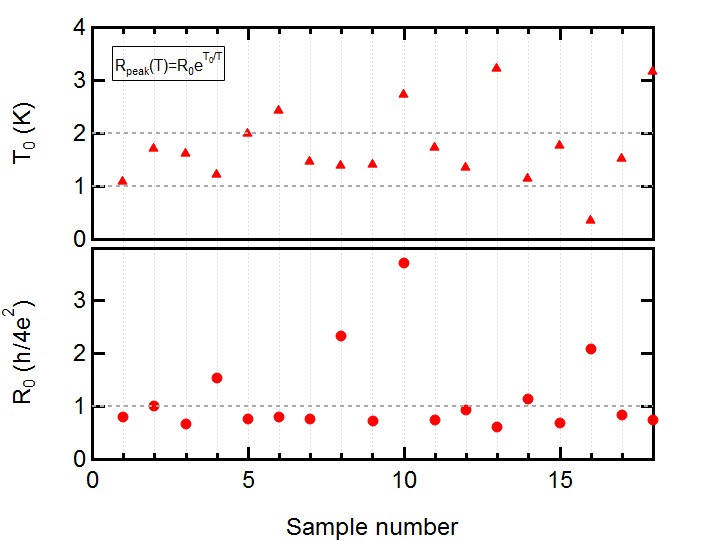}
		
		\caption{{\bf T$_{0}$ and R$_0$ for various samples.} 
			In 18 out of 23 samples we studied, the $R_{Peak}(T)$ fits an activated behavior. Displayed here are $T_{0}$ (top) and $R_{0}$ (bottom), extracted from the activated fit for these 18 samples. 
			For most samples $T_{0}\in (1,2)$K, which is the typical range of the superconducting $T_{c}$, and  $R_{0}$ is close to $\frac{h}{4e^2}$.
		}			
		\label{Figure3}
	\end{figure}

In activated transport the parameters $T_{0}$ and $R_{0}$ reflect the microscopic state of the conduction process. The fact that we can describe the $T$ evolution of $R_{peak}$ with single $T_0$ and $R_0$ indicates that the peak results from a unique microscopic state that evolves through different $B$'s. 
The values of the fit parameters, extracted from figure \ref{Figure2}, hold some information regarding the nature of this maximal $R$ state.
We note that the prefactor, $R_{0}$, is very close to the quantum of resistance for Cooper-pairs ($R_{Q}=\frac{h}{4e^2}$) and the activation $T$, $T_{0}$, is close to the superconducting $T_{c}$ at $B=0$ (1.5K for sample AD8a1mm, typically in our samples $T_{c}\in(1,2)K$).
In figure \ref{Figure3} we display $R_{0}/R_{Q}$ (circles, bottom graph) and $T_{0}$ (triangles, top graph) for 18 different samples we studied (see supplemental material section S3 for a table that includes the sample names and fit parameters values). 
It can be seen that for most samples $R_{0}/R_{Q}\in(0.6-1.15)$ and $T_{0}\in(1,2)$K. It is important to mention that, as discussed in ref \cite{murthyprl2004}, $T_{0}$ and the superconducting $T_{c}$ are of the same order but show opposite dependencies on disorder strength (sample number 16 has the highest $B_{c}$, therefore the lowest disorder strength and the lowest $T_{0}$ of 0.34K). 

We would like to stress that the data shown were extracted from 18 samples where the MR peak showed a good activation fit. The samples are very diverse, out of the 18 samples 4 were insulating at $B=0$, and 14 were superconducting with $B_{c}\in(0.1-4.3)$T. The area of the different samples varied from 50$\mu$m$^2$ to 2mm$^2$. We also studied 5 additional samples where the peak did not show a good activation fit. In these samples $R_{peak}$ had a reasonable variable range hopping behavior (see discussion section).
	
\textit{Discussion} - 

In figure \ref{Figure1}b,c we introduced two phenomenological fits that correspond to two possible $T=0$ phase diagrams. 
The main difference between these two fits is whether the $T\to 0$ limit of $B_{peak}$ is $B_{c}$ or some other $B>B_{c}$.
We would like to note that, as already discussed above, in the insulating phase close to the SIT, $R$ seems to diverge at a finite $T$ \cite{MaozFiniteT}, $T^{*}$, which depends on $B$, $T^{*}(B)$. 
As $T^{*}$ is finite over a continuous $B$-range, at non-zero $T$'s below the maximal $T^{*}$ $R$ will be infinite over a continuous $B$-range (assuming no other transport mechanism takes over at low $T$'s) and $B_{peak}$ will not be single valued. 
This picture will correspond to a third phase diagram where $B_{peak}^{0}$ spans over a $B$-range above $B_{c}$.

2-fluid model - 
One approach in theoretical and phenomenological modeling of the MR peak is to introduce two competing transport mechanisms, one dominant at low fields and high $T$'s and the other at high fields and low $T$'s \cite{galitski2005vortices,meirprb73}. 
Such models assume a continuous crossover between a low-$B$ and a high-$B$ transport regime. 
We attempted to reproduce our results using such a model.
The simplest phenomenological model we can create, adopting such a 2-fluid approach and without any microscopic assumptions, is the following: from the high-$T$ range ($0.2-1$ K), at various $B$'s, we extract the parameters $T_{0}^{ACT}(B)$ and $R_{0}^{ACT}(B)$, which are $T_{0}$ and $R_{0}$ of an activated fit to equation (\ref{eTdependence}), while from the low-$T$ range we extract an Efros-Skhlovskii $T^{ES}_{0}(B)$ and $R^{ES}_{0}(B)$, which are $T_{0}$ and $R_{0}$ of equation (\ref{eTdependence}) with $\gamma=0.5$ (variable range hopping dominates the transport at low $T$'s).

\begin{figure} [h!]
			\includegraphics [width=8.5 cm] {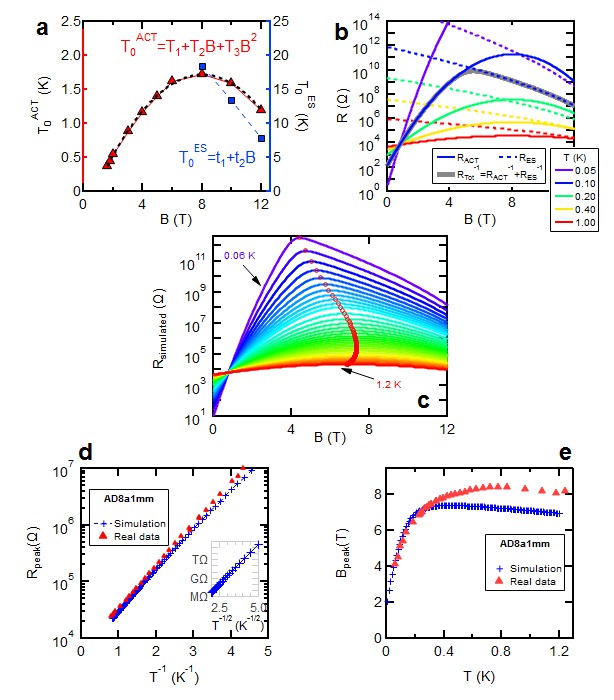}					
			\caption{{\bf Results of a 2-fluid model} 
			(a) $T_{0}^{ACT}$ (red, left axis) and $T_{0}^{ES}$ (blue, right axis) vs $B$. 
			(b) $R_{0}^{ACT}$ (continuous lines) and $R_{0}^{ES}$ (dashed lines) vs $B$. The color coding marks different $T$'s. The gray line is $R^{-1}=R_{ACT}^{-1}+R_{ES}^{-1}$ at $0.1$ K.
			(c) $R$ vs $B$ where the color coding marks different $T$'sand the MR peaks are marked by red circles. 
			(d) $R_{peak}$ (log scale) vs $T^{-1}$, the red triangles are the data displayed in \ref{Figure2}, the blue crosses are the simulated data using a 2-fluid model. Inset: the low-$T$ side of the simulated $R_{peak}$ vs $T^{-1/2}$ where the measured $R_{peak}$ becomes sub-activated. 
			(e) $B_{peak}$ vs $T$, the red triangles are the data displayed in \ref{Figure1}b, the blue crosses are from the 2-fluid model.
			}			
			\label{Figure4}
		\end{figure}

We then fit $T_{0}^{ACT}(B)$, $T^{ES}_{0}(B)$  (figure \ref{Figure4}a), $R_{0}^{ACT}(B)$, and $R^{ES}_{0}(B)$ to phenomenological functional forms (for simplicity we constrained both $R_{0}$'s to be constants) and extrapolate them to all $B$'s.  
In figure \ref{Figure4}b we plot the resulting $R_{ES}$ (dashed lines) and $R_{ACT}$ (continuous lines) vs $B$. 
If we assume that the activated and Efros-Skhlovskii behaviors extracted above are two parallel transport channels that are accessible at all $B$'s then $R^{-1}=R_{ACT}^{-1}+R_{ES}^{-1}$. The gray line in figure \ref{Figure4}b is the parallel $R$ at $0.1$ K. 
In figure \ref{Figure4}c we plot the resulting $R(B)$'s at different $T$'s where for each $T$ we mark the MR peak with a red circle.
From here one can qualitatively reproduce some of our main results.  

In figure \ref{Figure4}d we plot a comparison between $R_{peak}$ vs. $T^{-1}$ derived from this model (blue) and of our data (red). In the inset we display $R_{peak}$ vs $T^{-1/2}$ for the low-$T$'s where $R_{peak}$ is sub-activated.
In figure \ref{Figure4}e we plot $B_{peak}$ vs $T$ from this 2-fluid model (blue) and from the data displayed in figure \ref{Figure1}b (red). A benefit of such an approach is that it provides a way to break a complicated transport mechanism to its ingredients and study each of the two simpler, independent (by assumption), competing mechanisms separately.
We note that this model is lacking and does not explain the hyper-activated $R(T)$ observed below the MR peak \cite{MaozFiniteT}.

Non-activated behavior of $R_{peak}$ - 
Although the activation behavior of the peak was observed in the majority of samples we examined, there were also 5 samples (out of 23) were $R$ at the MR peak was sub-activated with a reasonable variable range hopping fit. 4 out of these 5 samples are in the high disorder limit and one in the opposite, low-disorder, limit. We did not manage to pin-point a criterion that makes these sample different. 
In addition, in some of the samples that showed an activated MR peak, there were deviations from activation at sufficiently low $T$'s.
One possible explanation can be that at low $T$'s, in the highly disordered samples, variable range hopping becomes beneficial and therefore dominates the low $T$ transport. This is demonstrated in the inset of figure \ref{Figure4}d where our model predicts a sub-activated peak that results from low $T$ variable range hopping.

The MR peak in other systems -  A MR peak is not unique to highly disordered superconductors and was also observed in various systems such as high-$T_{c}$ superconductors \cite{ando1995logarithmic, nakao1994magnetoresistance,yan1995negative}, Josephson junction arrays \cite{van1991phase}, granular superconductors \cite{gerber1997insulator}, quantum hall \cite{li1991low} and complex oxide interfaces \cite{caviglia2008electric}.
In some high $T_{c}$ superconductors a MR peak appears only below certain $T$'s, in addition and in contrast to our findings, in both $\rho_{ab}$ and $\rho_{c}$, $B_{peak}$ increases while cooling \cite{ando1995logarithmic, nakao1994magnetoresistance,yan1995negative}. 
Josephson junction arrays exhibit multiple MR peaks (and multiple SITs) \cite{van1991phase,delsing1998two} separated by minimas at $f=n/m$ where $f$ is the magnetic flux per cell in units of the magnetic flux quantum ($\Phi_{0}=h/2e$) and $n,m$ are integers. It seems that in these arrays $B_{peak}$ has no significant $T$ dependence.
A further comparison to other systems is beyond the scope of this work.

\begin{figure} [h!]
	\includegraphics [width=8.5 cm] {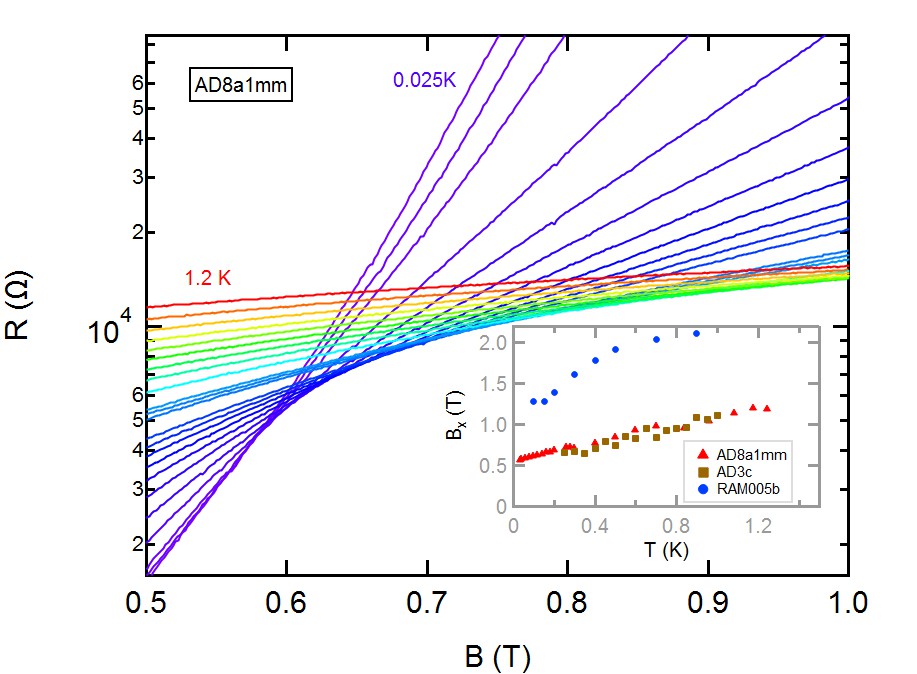}
	\caption{{\bf $T$ dependence of $B_{c}$} 
	$R$ (log scale) vs $B$ of sample AD8a1mm where we plot the same data of figure \ref{Figure1}a focusing on the crossing point. The color coding mark different isotherms. It can be seen that the crossing between two consecutive isotherms varies with $T$. Inset: $B_{x}$ vs $T$ for three different samples where the red triangles correspond to the main figure (AD8a1mm) and the blue and brown data are of the same samples as in figure \ref{Figure2}.
		}			
	\label{Figure5}
\end{figure}

$T$ dependence of $B_{c}$ - In figure \ref{Figure5} we re-plot the data of figure \ref{Figure1}a focusing on the vicinity of the crossing of different isotherms. 
Clearly there is no single $B_{c}$ where all isotherms coincide. Instead there are multiple crossing points and the crossing $B$ between successive isotherms ($B_{x}$) has a clear $T$ dependence. 
In the inset of figure \ref{Figure5} we plot $B_{x}$ vs $T$ for the same three samples displayed in figure \ref{Figure2} (see figure S4b of the supplementary material for $B_{x}(T)$ of additional samples). 

A $T$ dependence of $B_{x}$ was observed in several systems near their critical point. Near the SIT it was reported in highly disordered superconductors \cite{HebardPrl,steiner2005superconductivity} and high-$T_{c}$ superconductors \cite{ando1995logarithmic,nakao1994magnetoresistance} and near the superconductor-metal transition in complex oxide interfaces \cite{biscaras2013multiple} and ultra-thin Ga films \cite{xing2015quantum}.
As can be seen, in the high disorder limit of $\alpha$-InO films (samples with $B_{x}\ll H_{c2}\simeq 14$ T in figures \ref{Figure5} and S4b of the supplementary material and in reference \cite{HebardPrl}) $\frac{dB_{x}}{dT}>0$ and the $T\to 0$ limit of $B_{x}$ seems to be a finite positive value. 
On the other hand, in systems where superconductivity terminates with a metallic or weakly insulating phase such as $\alpha$-InO films in the clean limit (see two samples with $B_{x}\sim H_{c2}\simeq 14 $ T in figure S4b of the supplementary material and reference \cite{gantmakher1998resistance}), high-$T_{c}$ superconductors \cite{ando1995logarithmic,nakao1994magnetoresistance} and in some systems near the superconductor-metal transition \cite{biscaras2013multiple,xing2015quantum} the trend is opposite as $\frac{dB_{x}}{dT}<0$ (as expected if we replace $B_{x}$ with $H_{c2}$ \cite{tinkham}). 
We note that in our study and in the references where we managed to identify the $T$ dependences of both $B_{x}$ and $B_{peak}$ \cite{HebardPrl,ando1995logarithmic,nakao1994magnetoresistance} $\frac{dB_{x}}{dT}$ and $\frac{dB_{peak}}{dT}$ have the same sign \cite{Footnote2}.

The $T$ variation of $B_{x}$ poses difficulties in defining $B_{c}$ which results in several similar but non-equivalent definitions \cite{Footnote3}. 
In this work we defined $B_{c}$ as $B_{c}\equiv B_{x}(T\to0)$. For sample AD8a1mm this results in $B_{c}=0.55$ T.
 
In summary, we have showed that, as $T\to 0$, $B_{peak}$ decreases significantly which raises the interesting possibility that for sufficiently disordered samples $B_{peak}^{0}=B_{c}$.
We have showed that although the MR peak appears at different $B$'s, $R_{peak}$ is typically activated and can be described with single $R_{0}$ and $T_{0}$ suggesting that this maximally resistive state is a single microscopical state that emerges at lower and lower $B$'s.
The similarities of $R_{0}$ to $R_{Q}$ and of $T_{0}$ to the superconducting $T_{c}$ shows that this unique insulating state is tightly related to the microscopic superconducting nature of the insulating phase.


\begin{acknowledgments}
We are grateful to Y. Meir for fruitful discussions. This work was supported by the Israeli Science Foundation and the Minerva Foundation.
\end{acknowledgments}

\clearpage
\newpage
\setcounter{page}{1}
\renewcommand{\thepage}{\roman{page}}
\renewcommand{\thefigure}{S\arabic{figure}}
\renewcommand{\thesection}{S\arabic{section}}
\setcounter{figure}{0}
\setcounter{section}{0}

	\part*{\centering Supplemental material\\for\\The Temperature dependence of the magneto-resistance peak in highly disordered superconductors}

	\section{$T$-Activation fit}
\begin{figure} [h!]
		\includegraphics [width=8.5 cm] {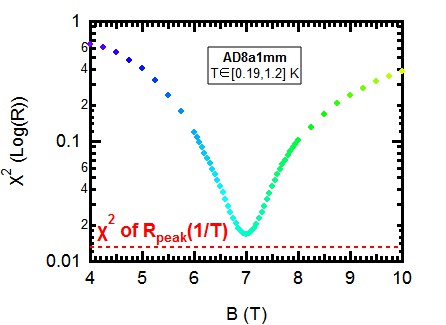}
		
		\caption{{\bf $\boldmath \chi^2$ test of T - Activation.} 
			$\chi^2$ of the fit $Ln(R)\propto T^{-1}$ at various $B$'s (dots) and at the MR peak (dashed red line).
		}			
		\label{FigureS1}
	\end{figure}

	In the main text we show that, although measured at different magnetic fields ($B$'s), the resistance ($R$) measured the magnetoresistance (MR) peak ($R_{peak}$) has an activated behavior that holds over a wide $T$ range. 
	In figure \ref{FigureS1} we display a $\chi^2$ test of the $T$-activation fit ($Ln(R)\propto T^{-1}$) of sample AD8a1mm for various constant $B$'s for $T\in (0.19,1.2)$K. 
	The dashed red line is $\chi^2$ of the $B$ dependent MR peak. The dots mark $\chi^2$ at each constant $B$. 
	It can be seen that $\chi^2$ at the peak is lower than at any constant $B$. This shows that the MR peak better fits an activation behavior than all constant $B$'s.
	
	\section{Algorithm for peak detection and error determination}
\begin{figure} [h!]
		\includegraphics [width=8.5 cm] {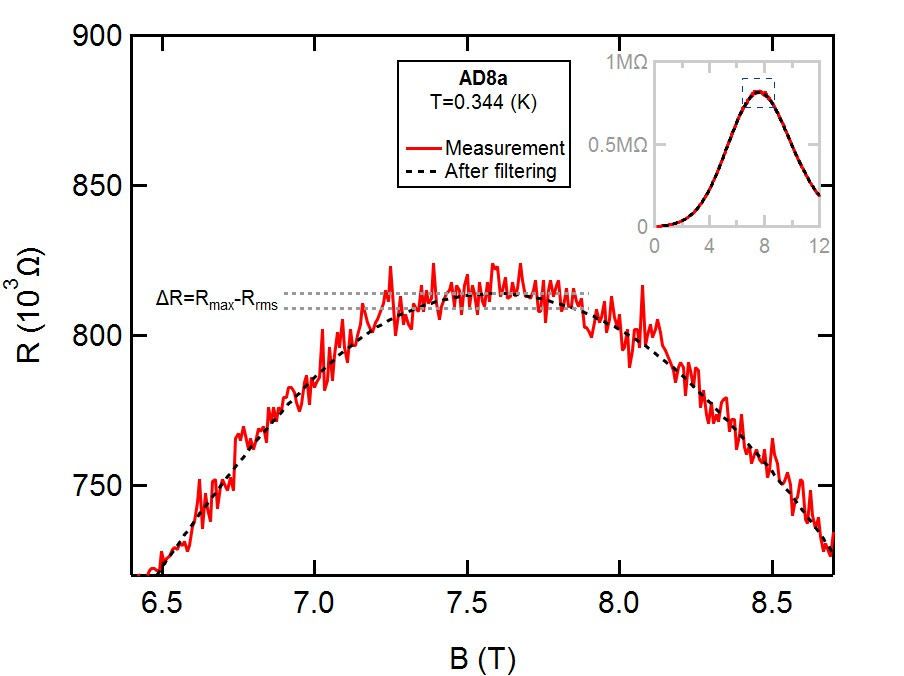}
		\includegraphics [width=8.5 cm] {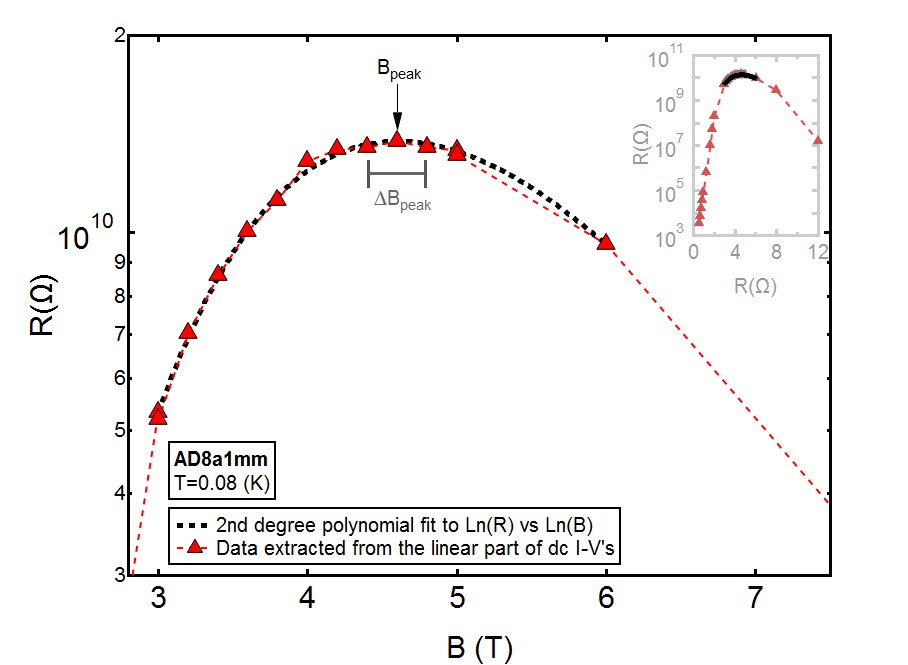}
		
		\caption{{\bf Peak detection and error determination} 
			$R$ vs $B$ (a) of a continuous (high $B$ resolution) measurement and (b) of a measurement with a large $B$ spacing (extracted from $I-V$'s measured at constant $B$'s). In the insets of both figures we display the full $B$ range of the measurements while the main figures focus on the peak. (a) The measured data is displayed in red, the dashed black line was extracted from the data using a numerical low-pass filter algorithm. (b) The measured data appears as red triangles. The dashed red line is a guide for the eye. The black line is a fit to the data around the peak. The details of how we determined $B_{peak}$ and it's error appears in the text.
		}			
		\label{PeakDetection}
	\end{figure}

	The values of $B_{peak}$ were derived from both continuous low-frequency 2 probe measurements and from isolated data points extracted from the linear portion of the dc current-voltage characteristics (used for $R>100$ M$\Omega$). 
	Due to the large difference in the density of data points near the peak, for the different measurement methods we used different methods for extracting $B_{peak}$.
	In figure \ref{PeakDetection}a we illustrate the method we used to determined the value and error bars of $B_{peak}$ from continuous data.
	
	In the figure we display $R$ vs $B$ for sample AD8a1mm at $T=0.344$K (red data) where in the inset we display the full $B$-scale of the measurement and in the main figure we zoom into the vicinity of the MR peak. 
	The dashed black line was extracted from the (measured) red data using a numerical low-pass filter algorithm. 
	$B_{peak}$ was determined as $B$ where the maximum of the filtered results appears (the filtered results have a single, well-defined, peak).
	The error bars were determined in three steps: 1. We calculated the $R$ measurement uncertainty, $R_{rms}$ as maximal value of $R_{measured}-R_{filtered}$ near the peak. 2. We defined a $R$ uncertainty interval, $\Delta R=R_{peak}-R_{rms}$ (as appears in the figure). 3. We found the $B$ values where $R_{filtered}=R_{peak}-\Delta R$ and extended the error bars from $B_{peak}$ to $1/\sqrt{(2)}$ of the $B$ distance to these values.
	
	In figure \ref{PeakDetection}b we display $R$ (log scale) vs $B$ of the less-dense (in $B$) data and illustrate the method used to determine $B_{peak}$.
	The red triangles were extracted from the Ohmic regime of a dc current-voltage characteristics. 
	The black line is a fit to the data points surrounding the maximal measured $R$. The fit function we used is of a phenomenological functional form - a second degree polynomial of $Ln(R)$ in $Ln(B)$.
	We chose $Ln(B)$ for the abscissa because the increase in $R$ below the peak is faster than it's decrease above the peak. A choice of $B$ instead of $Ln(B)$ will not change any of the qualitative results.
	$B_{peak}$ was defined by the maximum of the fit. The error bars we set as the $B$ measuring resolution around $B_{peak}$.

	\section{Activation behavior, fit parameters and $B_{peak}(T)$ for different samples}
\begin{figure} [h!]
		\includegraphics [width=8.5 cm] {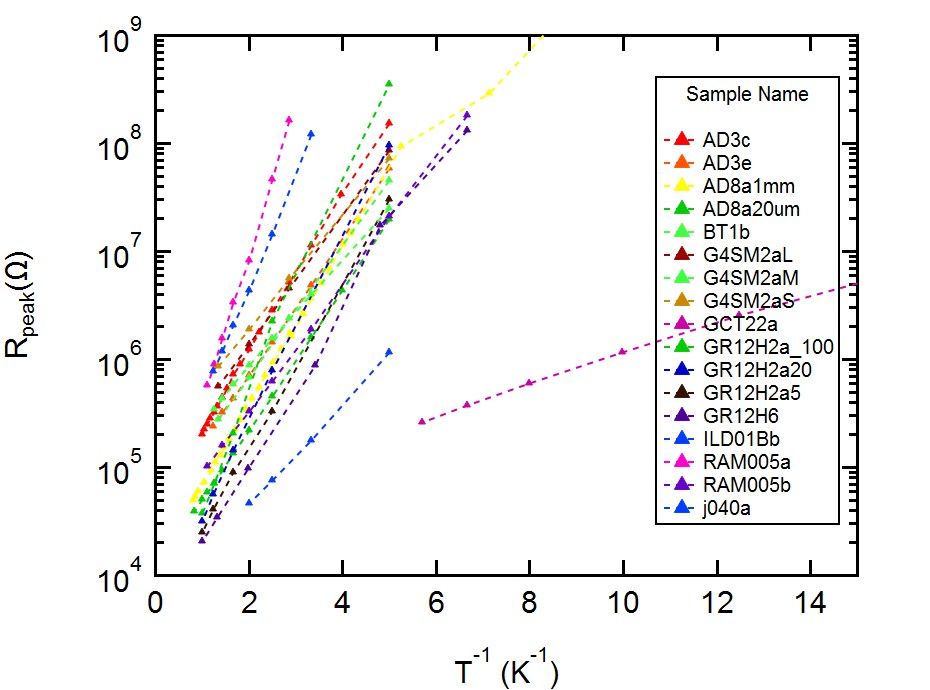}
		
		\caption{{\bf Samples with an activated peak} 
		}			
		\label{FigRpeakvsinvT}
	\end{figure}

	In figure \ref{FigRpeakvsinvT} we display $R_{peak}$ (log scale) vs $T^{-1}$ for the 18 samples where the peak showed a reasonable $T$ activation behavior.
	
	The activation fit parameters appear in table \ref{Table1} where the number in the left column refers to the "sample number" abscissa in figure 3 of the main text, $T_{0}$ and  $R_{0}$ were extracted from the fit $R_{peak}(T)=R_{0}exp(T_{0}/T)$.
	
	\begin{table} [h!]
		\centering
		\begin{tabular}{|c | c | c | c|} 
			\hline
			Number & Sample name & $T_{0}$ [K] & $R_{0}$ [$R_{Q}$]\\ [0.5ex] 
			\hline\hline
			1 & ILD01Bb & 1.08 & 0.79 \\ 
			\hline
			2 & AD8a1mm & 1.7 & 0.99 \\
			\hline
			3 & GR12H6 & 1.61 & 0.65 \\
			\hline
			4 & G4SM2aS & 1.21 & 1.53 \\
			\hline
			5 & GR12H2a20 & 1.99 & 0.75 \\
			\hline
			6 & j040a & 2.4 & 0.79 \\ 
			\hline
			7 & AD3e & 1.46 & 0.75 \\
			\hline
			8 & G4SM2aL & 1.38 & 2.32 \\
			\hline
			9 & G4SM2aM & 1.4 & 0.72 \\ 
			\hline
			10 & GR12H2a100 & 2.72 & 3.7 \\
			\hline
			11 & AD3c & 1.72 & 0.73 \\
			\hline
			12 & RAM005b & 1.34 & 0.93 \\
			\hline
			13 & RAM005a & 3.21 & 0.61 \\ 
			\hline
			14 & BT1b & 1.14 & 1.13 \\
			\hline
			15 & GR12H2a5 & 1.76 & 0.68 \\
			\hline
			16 & GCT22a & 0.34 & 2.08 \\	 
			\hline
			17 & AD8a20um & 1.51 & 0.83 \\
			\hline	 
			18 & AD1a & 3.16 & 0.74 \\ [1ex] 
			\hline	 	 
		\end{tabular}
		\caption{$T$-activation Parameters of the MR peak}
		\label{Table1}
	\end{table}
	
\begin{figure} [h!]
		\includegraphics [width=8.5 cm] {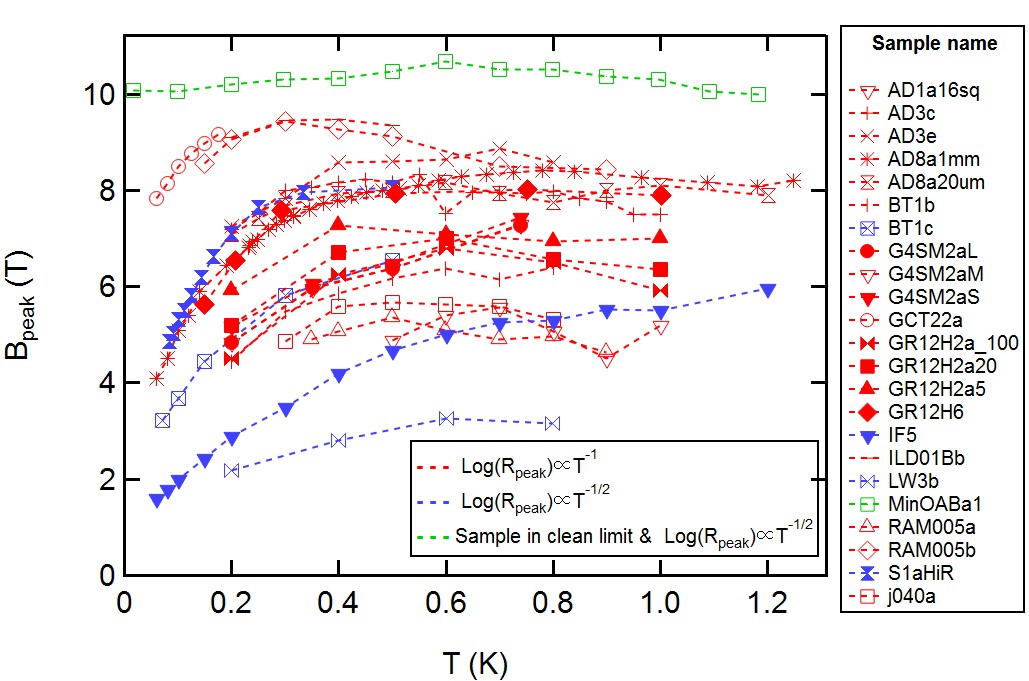}
		\includegraphics [width=8.5 cm] {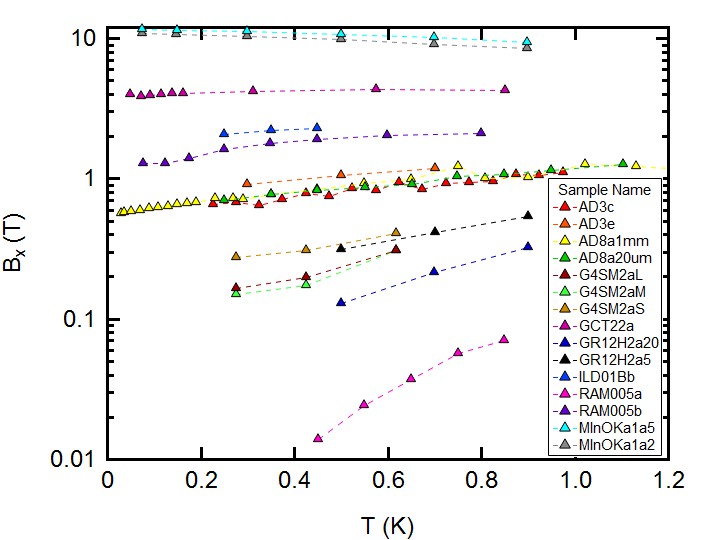}
		
		\caption{{\bf $B_{peak}$ and $B_{x}$ vs $T$ for various samples}
		}			
		\label{FigBpeakvsT}
	\end{figure}

	In figure \ref{FigBpeakvsT} we display $B_{peak}$ vs $T$ of all 23 samples examined in this work. 
	The red data points correspond to samples with an activated MR peak, the blue data corresponds to highly disordered samples where $R_{peak}$ has a reasonable variable-range-hopping fit, the green data corresponds to a sample in the clean limit that also fit variable-range-hopping.
	
	\section{Simulation fits}
	In the discussion section of the main-text we present a 2-fluid simulation where we assume that the $B$ dependence of the $R$ is a result of two independent competing transport mechanisms, one dominant at high $B$'s and the other at low $B$'s.
	To gain information regarding the different "fluids" we used the simplest phenomenological form we thought of and followed the following procedure; First we assumed that far below the MR peak, in the insulating phase, $R_{activation}(T)=R_{0}exp(T_{0}/T)$. We then extracted the activation fit parameters $R_{0}$ and $T_{0}$ (see table \ref{Table2}) for various constant $B$'s on both sides of the MR peak (the fit focused mainly on the higher $T$'s where the data is indeed activated). 
	We repeated the same procedure for $B$'s far above the MR peak where we fit the data of constant $B$'s to an Efros-Shklovskii (ES) variable-range-hopping behavior, $R_{ES}(T)=R_{ES,0}exp((T_{ES,0}/T)^{1/2})$, and extracted $R_{ES,0}$ and $R_{ES,0}$ (see table \ref{Table2}, the fit focused mainly on the low-$T$'s where variable-range-hopping should dominate).
	\begin{table} [h!]
		\centering
		\begin{tabular}{|c | c | c | c| c|} 
			\hline
			$B$ [T] & $R_{0}^{ACT}$ [$\Omega]$ & $T_{0}^{ACT}$ [K] & $R_{0}^{ES}$ [$\Omega]$ & $T_{0}^{ES}$ [K] \\ [0.5ex] 
			\hline\hline
			1.6	&	7111	&	0.372	&	-	&	-	\\
			\hline									
			1.8	&	7309	&	0.448	&	-	&	-	\\
			\hline									
			2	&	7093	&	0.548	&	-	&	-	\\
			\hline									
			3	&	7382	&	0.88	&	-	&	-	\\
			\hline									
			4	&	7136	&	1.166	&	-	&	-	\\
			\hline									
			5	&	6680	&	1.403	&	-	&	-	\\
			\hline									
			6	&	5943	&	1.62	&	-	&	-	\\
			\hline									
			8	&	6125	&	1.715	&	1580	&	18.443	\\
			\hline									
			10	&	6592	&	1.585	&	1580	&	13.423	\\
			\hline									
			12	&	8052	&	1.196	&	1580	&	7.752	\\
			\hline									
		\end{tabular}
		\caption{Activation and ES fit parameters of sample AD8a1mm at various $B$'s}
		\label{Table2}
	\end{table}
	
	In order to extend the activated conductivity channel to the whole $B$ range we used the following phenomenological fits for the activation parameters $R_{0}(B)=r_{0}$, $T_{0}(B)=t_{0}+t_{1}B+t_{2}B^{2}$ (we had to take the second order in $B$ to account for the peak in $T_{0}(B)$) where $r_{0}=6.942 k\Omega$, $t_{0}=-0.389$ K, $t_{1}=0.525$ K/T and $t_{2}=-0.033$ K/T$^{2}$.
	In order to extend the ES conductivity channel to the whole $B$ range we used the following phenomenological fits for the ES parameters $R_{ES,0}(B)=r_{a}$, $T_{ES,0}(B)=t_{a}+t_{b}B$ where $r_{a}=1.58 k\Omega$, $t_{a}=40$ K and $t_{b}=-2.67$ K/T.
	Using these fits we get a complete functional fit of $R(B,T)$ from which we calculated the $B_{peak}(T)$ and $R_{peak}(T)$ displayed in figure 4 of the main-text.
	

\begin{thebibliography}{1}
\bibitem[{Foo({\natexlab{a}})}]{Footnote1}
	\bibinfo{note}{As at $T=0$ the resistance in the insulating phase is infinite,
		the MR peak is not well defined at $T=0$. Therefore one should first consider
		$B$ of the MR peak at a finite $T$ and then take the limit $T\to0$}.
\end{thebibliography}
\end{document}